\documentclass[useAMS,usegraphicx]{mn2e}
\usepackage{graphicx}
\usepackage{color}

\begin{document}
\newcommand{\be}{\begin{equation}}
\newcommand{\ee}{\end{equation}}
\newcommand{\ba}{\begin{eqnarray}}
\newcommand{\ea}{\end{eqnarray}}
\def\apj {ApJ}
\def\aj {AJ}
\def\aa {A \& A}
\def\mnras {MNRAS}
\newcommand{\mincir}{\raise
-2.truept\hbox{\rlap{\hbox{$\sim$}}\raise5.truept\hbox{$<$}\ }}
\newcommand{\magcir}{\raise
-2.truept\hbox{\rlap{\hbox{$\sim$}}\raise5.truept\hbox{$>$}\ }}

\title{A Consistent Comparison of Bias Models using Observational Data}

\author[A. Papageorgiou et al.]
{
  \parbox[t]{\textwidth}
  {
  A. Papageorgiou${^{1,2}}$, M. Plionis${^{1,3}}$, S. Basilakos${^{4,5}}$, C. Ragone-Figueroa$^6$
    }
  \vspace*{6pt}\\ 
  \parbox[t]{15 cm}
  {
    $^1$ Institute of Astronomy \& Astrophysics, National Observatory of Athens,
    Palaia Penteli 152 36, Athens, Greece.\\
    $^2$Faculty of Physics, Department of Astrophysics, Astronomy \&
    Mechanics University of Athens, Panepistemiopolis, Athens 157 83,
    Greece \\
    $^3$ Instituto Nacional de Astrof\'{\i}sica Optica y Electr\'onica, AP 51
    y 216, 72000, Puebla, M\'exico.\\
    $^4$Academy of Athens, Research Center for Astronomy \& Applied Mathematics,
 Soranou Efesiou 4, 11527, Athens, Greece \\
    $^5$ High Energy Physics Group, Dept. ECM, Universitat de Barcelona,
Av. Diagonal 647, E-08028 Barcelona, Spain  \\
    $^6$ Instituto de Astronom\'ia Te\'orica y Experimental, IATE,
    CONICET-Observatorio Astron\'omico, Universidad Nacional de
    C\'ordoba, Laprida 854, X5000BGR, C\'ordoba,
Argentina
  }
}
\date{\today}

\maketitle

\begin{abstract}

\indent
We investigate five different models for the dark matter halo bias,
ie., the ratio of the fluctuations of mass tracers
to those of the underlying mass, by comparing their cosmological
evolution using optical QSO and galaxy bias data at different
redshifts, consistently scaled to the WMAP7 cosmology.
Under the assumption that each halo hosts one extragalactic mass
tracer, we use a $\chi^2$ minimization procedure to determine the free
parameters of the bias models as well as to statistically
quantify their ability to represent the observational
data. Using the Akaike information criterion 
we find that the model that represents best the observational data is the
Basilakos \& Plionis (2001; 2003) model with the tracer merger
extension of Basilakos, Plionis \& Ragone-Figueroa (2008) model. 
The only other statistically equivalent model, as indicated by
the same criterion, is the Tinker et al. (2010) model.
Finally, we find an average, over the different models, 
dark matter halo mass that hosts optical QSOs of:
$M_h\simeq 2.7 (\pm 0.6) \times 10^{12} h^{-1} M_{\odot}$, while
the corresponding value for optical galaxies is:
$M_h\simeq 6.3 (\pm 2.1) \times 10^{11} h^{-1} M_{\odot}$.
\end{abstract}

\section{Introduction}
\indent
It is of paramount importance for cosmological and galaxy formation
studies the understanding of how galaxies and other extragalactic
mass-tracers relate to the underlying distribution of matter.  
The current galaxy formation paradigm assumes that galaxies form
within dark matter haloes, identified as high-peaks of an underlying
initially Gaussian density fluctuation field, and that they trace in a
biased manner such a field (eg., Kaiser 1984; Bardeen et al. 1986).
A formation process of this sort can explain the difference of the clustering
amplitude between the different extragalactic mass tracers (galaxies,
groups and clusters of galaxies, AGN, etc) as being due to the
different bias among the underlying density field and that of the dark
matter (DM) halos that host the mass tracers. 

In order to quantify such a difference,
one can use the so-called linear
bias parameter $b$, which for continuous density fields
is defined as the ratio of the fluctuations of the mass tracer
($\delta_{\rm tr}$) to those of the underlying mass ($\delta_{m}$):
\be
b=\frac{\delta_{\rm tr}}{\delta_{m}} \;,
\ee	
Based on this definition one can write the bias parameter in 
a number of equivalent ways: (a) as the square root of the ratio of
the two-point correlation function of the tracers to the underlying mass:
\be
\label{bias1}		
b=\left(\frac{\xi_{\rm tr}}{\xi_{m}}\right)^{1/2}\;,
\ee
since $\xi(r)=\langle \delta({\bf x}) \delta({\bf x}+{\bf r})\rangle$,
in which case one considers the large-scale correlation function (ie.,
scales $\magcir 1 h^{-1}$ Mpc), corresponding roughly to the so-called
halo-halo term of the DM halo correlation function
(eg., Hamana, Yoshida, Suto 2002),
and (b) as the ratio of the variances of the tracer and underlying mass
density fields, smoothed at some linear scale traditionally 
taken to be $8 \; h^{-1}$ Mpc (at which scale the variance is of
order unity):
\be
\label{bias2}		
b=\frac{\sigma_{8,\rm tr}}{\sigma_{8, m}}\;,
\ee
since $\sigma^2_{8}=\xi(0)=\langle \delta^2({\bf x})\rangle$.

A further important ingredient in theories of structure formation, is
the cosmological evolution of the DM halo bias parameter (eg., Mo \& White
1996; Tegmark \& Peebles 1998, etc). 
A large number of such bias evolution models have been presented in
the literature and the aim of this work is to compare them
using as a criterion how well do they fit the observed bias, at different redshifts, of
optical QSOs and galaxies. In such a comparison we will make the
simplified assumption that each DM halo hosts one mass tracer. This is
consistent with the definition of the linear bias, where one uses
either the large-scale correlation function (which corresponds to
the halo-halo term) or the smoothed
to linear scales variance of the fluctuation field, while any residual
non-linearities will probably be suppressed in the ratio of the tracer to
underlying mass correlation functions or variances.
Further suppression of non-linearities, introduced for example by redshift-space
distortions, can be achived using the integrated correlation
function within some spatial scale; see discussion in section 2 below.
 
There are two basic families of analytic bias evolution models. The
first, called the  {\em galaxy merging} bias model, utilizes the halo mass
function and is based on the Press-Schechter (1974) formalism, 
the peak-background split (Bardeen et al. 1986) and
the spherical collapse model (Cole \& Kaiser 1989; Mo \& White 1996, Matarrese et
al. 1997; Moscardini et al. 1998; Sheth \& Tormen 1999; Valageas 2009; 2011).
Cole \& Kaiser (1989) found for the bias evolution that 
$$b(M,z)= 1/(1+z)-1/\left[1.68(1+z)\right]+1.68(1+z)/\sigma^2(M)\;,$$ 
where $\sigma^2(M)$ is the variance of the mass fluctuation field,
while Mo \& White (1996) derived for an Einstein-de Sitter universe that
$$b(z)=0.41+\left[b(0)-0.41\right] (1+z)^{2}\;.$$ 
Mo, Jing \& White (1997)
extended the previous study in the quasi-linear regime by taking
into account high order correlations of peaks and halos. 
Similarly, Matarrese et. al. (1997) estimated the bias in a merging
model where the halo mass exceeds a certain threshold. They found that for an
Einstein-de Sitter universe:
$$b(z)=0.41+\left[b(0)-0.41\right] (1+z)^{\beta}\;,$$ 
while Moscardini et. al. (1998) 
generalized the above bias evolution model for a variety cosmological
models. 

Many studies have compared the prediction of the {\em merging}
bias model with numerical simulations 
and beyond an overall good agreement, differences have been found in the
details of the halo bias. 
For example, the spherical collapse
model under-predicts the halo bias for low mass halos and fails to
reproduce the dark matter halo mass function found in simulations.
To solve this problem, Sheth, Mo \& Tormen (2001) extended their 
original model to include the effects of ellipsoidal collapse.
However according to
Tinker et. al. (2010), this model under-predicts the
clustering of high-peaks halos while over-predicts the bias of low mass
objects. Furthermore, Manera et al. (2009) and Manera \& Gaztanaga
(2011) find that the clustering of massive halos cannot be reproduced
from their bias calculated using the peak-background split.

Such and other differences have lead to other modifications of the models,
either suggesting new fitting bias
model parameters (eg., Jing 1998; Tinker et al. 2005), or new forms
of the bias model fitting function (eg. Seljak \& Warren 2004; 
Pillepich et al. 2010; Tinker et al. 2010) or even a non-Markovian
extension of the excursion set theory (Ma et al. 2011). 
A further step was provided by de Simone, Maggiore \& Riotto (2011), 
who incorporated the effects of ellipsoidal collapse to the original
Ma et al. model, which is based on spherical collapse.

The second family of bias evolution models assumes a continuous
mass-tracer fluctuation field, proportional to that of the underlying
mass, and the tracers act as ``test particles''. In this context, the
hydrodynamic equations of motion and linear perturbation theory are 
applied. This family of models can be divided into two sub-families: 

\noindent
(a) The so-called {\em galaxy
  conserving} bias model uses the continuity equation and
the assumption that tracers and underlying mass share the same velocity
field (Nusser \& Davis 1994; Fry 1996; Tegmark \& Peebles 1998; Hui \&
Parfey 2007; Schaefer, Douspis \& Aghanim 2009).
Then the bias evolution is given as the solution of a 1st order
differential equation, and
Tegmark \& Peebles (1998) derived: 
$$b(z)=1+[b(0)-1]/D(z)\;,$$ 
where $b(0)$ is the bias factor at the present time and $D(z)$ the
growing mode of density perturbations. However, this bias model suffers from 
two fundamental problems: {\it the unbiased problem} ie., the fact that an unbiased set of
tracers at the current epoch remains always unbiased in the past, 
and {\it the low redshift problem} ie., the fact that this model
represents correctly the bias evolution only at relatively low
redshifts $z\mincir 0.5$ (Bagla 1998). Note that Simon (2005) has
extended this model to also include an evolving mass tracer population
in a $\Lambda$CDM cosmology.

\noindent
(b) An extension of the previous model, based on the basic
differential equation for the evolution of linear density perturbations, which 
implicitly uses that mass tracers and underlying mass share 
the same gravity field, and on the assumptions of linear and
scale-independent bias, provides a second order differential equation
for the bias. Its approximate solution provides the functional form
for the cosmological evolution of bias (Basilakos \& Plionis 2001;
2003 and Basilakos, Plionis \& Ragone-Figueroa 2008; hereafter BPR
model). The provided solution applies to
cosmological models, within the framework of general relativity, with
a dark energy equation of state parameter being independent of cosmic time (ie.,
quintessence or phantom). An extension of this model to engulf also
time-dependent dark energy equation of state models, including
modified gravity models (geometric dark energy), 
was recently presented in Basilakos, Plionis \& Pouri (2011).

The outline of this paper is as follows. 
In section 2 we present the data that we will use, we review the basic
techniques used in measuring the bias from samples of extragalactic 
objects and we will present the rescaling method used
in order to transform different bias data to the same (WMAP7) cosmology
(ie., flat $\Lambda$CDM with $\Omega_m=0.273$ and $\sigma_8=0.81$).
In section 3 we introduce the
different bias evolution theoretical models that 
we will investigate, while in section 4
we present our results and discussion. The main conclusions are
presented in section 5. In the Appendix we discuss the simulations
used to fit the free parameters of the BPR model, as well as the 
cosmological dependence of these parameters.

\section{Mass Tracer Bias Data}
\indent

The mass tracers that we will use in this work are optical QSOs
and galaxies, for which their linear bias with respect to the underlying mass
is available as a function of redshift. 
In particular, we will use:
\begin{itemize} 
\item The 2dF-based QSO results of Croom et
al. (2005), which are based on spectroscopic data of over 20000 QSOs 
covering the redshift range $0.3\leq z\leq2.2$ and on a 
$\Lambda$CDM cosmology with $\Omega_m=0.27$ and $\sigma_8=0.84$.
\item The SDSS (DR5) QSO ($z\mincir 2.2$) results of Ross et
al.(2009) based on spectroscopic data of $\sim$30000 QSOs and and on a 
$\Lambda$CDM cosmology with $\Omega_m=0.237$ and $\sigma_8=0.756$.
\item the SDSS (DR5) QSO results of Shen et
al. (2009), who used a homogeneous sample of $\sim$38000 QSOs within 
$0.1\leq z\leq5$ and on a $\Lambda$CDM cosmology with $\Omega_m=0.26$
and $\sigma_8=0.78$. In this case we will use only their $z\magcir
2.2$ results to avoid including in our analysis correlated
measurements of the bias, for the redshift range covered also by the
Ross et al. analysis.
\end{itemize}
Although there are other QSO bias data available, like the Myers
et. al. (2006) analysis of 300000 photometrically classified
SDSS DR4 QSOs, within $0.75\leq z\leq2.8$, we do not include
them in our analysis in order to avoid, in the redshift range studied,
as much as possible correlated bias measurements.

As far as galaxy data are concerned, we will use the bias results of 
Marinoni et. al. (2005), which are based on 3448 galaxies from the VIMOS-VLT
Deep Survey (VVDS), cover the redshift range: $0.4\leq z\leq1.5$
and use a $\Lambda$CDM cosmology with $\Omega_m=0.3$
and $\sigma_8=0.9$. 

Although in the next section we sketch the usual procedures used to
estimate the linear bias of a sample of extragalactic mass tracers, we
would like to stress that for the QSO data used in this work, the corresponding authors,
in order to minimize non-linear effects,
have estimated the integrated correlation function for scales $>1 h^{-1}$
Mpc, which in the usual jargon corresponds roughly to the halo-halo
term of the DM halo correlation function. As
for the VVTS galaxy bias data, Marinoni et al., devised a
procedure to estimate the bias of a smooth galaxy density field in
pencil beam surveys, disentangling the non-linear effects, and thus the
bias values used in this work correspond to the linear bias.

\subsection{Estimating the tracer bias at different redshifts}
Although we will use the bias data provided by the previous
references, for completion we briefly present here the basic methodology 
used to estimate the bias of some extragalactic mass tracer at a
redshift interval $z\pm \delta z$, using any of the basic
definitions of eq.(1)-(3).

The first issue that one has to keep in mind is that what we measure
from redshift catalogues is the redshift-space distorted value of
either the tracer correlation function, $\xi_{\rm tr}(s)$, or the
variance of the tracer density field $\sigma^2_{8, {\rm tr}, s}$ 
(the index $s$ indicates
redshift-space distorted spatial separations, while the index $r$
indicates true spatial separations). One needs to correct
for such distortions, resulting from the peculiar velocities of the
mass tracers, in order to recover the true spatial value of either
measures. Kaiser (1987) provides such a correction procedure which
entails in dividing the directly measured from the data tracer
correlation function or variance with a function
$F(\Omega_m,\Omega_\Lambda,b,z)$, given by (see also Hamilton 1998 and
Marinoni et al. 2005):
\be
\label{xi(s)}
F(\Omega_m,\Omega_\Lambda,b,z)= 1+\frac{2}{3}\beta(z)+\frac{1}{5}\beta^2(z)
\ee
with $\beta(z)=\Omega_m^{\gamma}(z)/b(z)$, and $\gamma=6/11$ for the
$\Lambda$CDM (eg., Wang \& Steinhardt 1998; Linder 2005), 
which implies that $\beta(z)=\Omega^{6/11}_mE(z)^{-12/11}(1+z)^{18/11}/b(z)$.
Therefore the relation between the redshift-space and real-space
measures used to estimate the bias parameter is:
\be\label{eq:red-real}
\frac{\xi_{\rm tr}(s,z)}{\xi_{\rm tr}(r,z)}=\frac{\sigma^2_{8, 
{\rm tr},s}(z)}{\sigma^2_{8, {\rm tr},r}(z)}= F(\Omega_m,\Omega_\Lambda,b,z)
\ee

Then combining equations (\ref{bias1}) or (\ref{bias2}) with 
(\ref{xi(s)}) and (\ref{eq:red-real})
provides the real-space bias factor according to:
\begin{eqnarray}\label{eq:final_bias}
b(z) =  \left[ \frac{\xi_{\rm tr}(s,z)}{\xi_{m}(r,z)}-
\frac{4\Omega_m^{12/11}(z)}{45}\right]^{1/2}
-\frac{\Omega_m^{6/11}(z)}{3} \nonumber \\
 = \left[ \frac{\sigma^2_{8, {\rm tr},s}(z)}{\sigma^2_{8, m}(z)}-
\frac{4\Omega_m^{12/11}(z)}{45}\right]^{1/2}
-\frac{\Omega_m^{6/11}(z)}{3}
\end{eqnarray}
where $\xi_{m}(r)$ and $\sigma^2_{8,m}$ are the
corresponding correlation function and variance of the underlying dark
matter distribution, given by the Fourier transform of the 
spatial power spectrum $P(k)$ of the matter fluctuations, linearly
extrapolated to the present epoch: 
\be
\xi_{m}(r, z)=\frac{D^2(z)}{2\pi^{2}}
\int_{0}^{\infty} k^{2}P(k) 
\frac{{\rm sin}(kr)}{kr}{\rm d}k \;\;,
\ee
and
\be
\label{eq:spat1}
\sigma^2_{8, m}(z)=\frac{D^2(z)}{2\pi^2}\int^{\infty}_{0}k^2P(k)W^2(kR_8)dk\;\;,
\ee
with $D(z)$ the normalized perturbation's growing mode (ie., such
that $D(0)=1$), $P(k)$ the CDM power spectrum given by: 
\be\label{eq:PS}
P(k)=P_{0} k^{n}T^{2}(\Omega_m,k)\;, 
\ee
with $T(\Omega_m, k)$ being the CDM transfer function 
(Bardeen et al. 1986; Sugiyama 1995; Eisentein \& Hu 1998), $n$ the slope of the
primordial power-spectrum (which according to WMAP7 is $=0.967$)
and $W(k R_8)$ the Fourier transform
of the top-hat smoothing kernel of radius $R=R_8=8 h^{-1}$ Mpc, given by
$W(kR_{8})=3({\rm sin}kR_{8}-kR_{8}{\rm cos}kR_{8})/(kR_{8})^{3}$.  

Now, although in the case of using eq.(3), the $\sigma_8$ variance is
free of non-linear effects by definition, this is not so when using the correlation
function approach (eq. 2). Therefore, in order to minimize nonlinear effects at
small separations one can replace $\xi_{\rm tr}(s)$ in eq.(\ref{eq:final_bias})
with the integrated correlation function, $\bar{\xi}_{\rm tr}(s)$.

An alternative approach in order to avoid redshift-space distortions is
to resolve the redshift-space separation, $s$, into two
components, one perpendicular ($r_p$) and one 
parallel ($\pi$) to the line-of-sight (see Davis \& Peebles 1983) 
and then estimating the 2-point
projected correlation function $w_p(r_p)$ along the perpendicular
dimension (within some range of the parallel dimension, say
$\pi_{min}<\pi<\pi_{max}$), which is related to the spatial correlation
function, $\xi(r)$, according to:
\be
w_p(r_p)=\int^{\pi_{max}}_{\pi_{min}}\xi(r_p,\pi)d \pi =
\int^{\pi_{max}}_{\pi_{min}}\frac{r\xi(r)}{\sqrt{r^2-r^2_p}}dr
\ee
where $\pi=|\Delta d|$ and $r_p= \Delta d \tan
\theta/2$, with $\Delta d$ the radial comoving distance separation of
any pair of mass tracers and $\theta$ is angular separation on the sky
of the pair members. 
As before, one can use the integrated correlation function, $\bar{\xi}$,
in order to minimize nonlinear effects.

Additionally, one can also use the angular two-point correlation
function, $w(\theta)$, instead of $\xi(s)$ or $w_p(r_p)$, in order to
obtain $\xi(r)$ via Limber's inversion, a procedure which 
also avoids the peculiar velocity distortions, but is hampered by the
necessity of {\em a priori} knowing the redshift selection function of
the mass tracers.

\subsection{Scaling the bias data to the same Cosmology}
\indent

Since different authors have estimated the optical QSO and galaxy bias
using different cosmologies,
 we need to convert them to the same cosmological background in order to be able to
use them consistently. As such we choose the recent
WMAP7 cosmology (Komatsu et al. 2011).

The procedure that we will follow uses the different $\sigma_8$
power-spectrum normalizations (eq. 3).
We wish to translate the value of bias from one cosmological model,
say $B$, to another, say $A$. The definition of bias at a redshift $z$ for these two
different cosmologies are given by:
\be
\label{ff1}
b_A(z)=\frac{\sigma_{8,{\rm tr},r,A}(z)}{\sigma_{8,m,A}(z)}
\ee
and 
\be
\label{ff2}
b_B(z)=\frac{\sigma_{8,{\rm tr},r,B}(z)}{\sigma_{8,m,B}(z)}
\ee
where the numerator is the real space value of $\sigma_8(z)$ estimated
directly from the data, using also eq.(\ref{eq:red-real}) to correct for
redshift space distortions. 
Dividing now  equation (\ref{ff1}) by (\ref{ff2}), taking
into account eq.(\ref{eq:red-real}), and making the fair assumption that:
\be
\sigma_{8,{\rm tr},s,A}(z)\simeq \sigma_{8,{\rm tr},s,B}(z)\;,
\ee 
since the different cosmologies enter only weakly in the
observational determination of $\sigma_{8,\rm tr}$, 
through the definition of distances, we then have:
\be
\label{ff3}
b_A(z)\simeq b_B(z)\frac{\sigma_{8,m,B}(z)}{\sigma_{8,m,A}(z)}
\left[\frac{F(\Omega_{m,B},\Omega_{\Lambda,B},b_B,z)}
{F(\Omega_{m,A},\Omega_{\Lambda,A},b_A,z)}\right]^{1/2}\;.
\ee
As it can be realized the required rescaled real-space bias, $b_A$, enters also in the
right hand side of the above equation, making it rather complicated to
analytically derive the full expression (using for example
eq.\ref{eq:final_bias}). However, noting that the redshift-space
distortion correction enters in the
scaling of the bias, from one cosmology to another, as the
square-root of the ratio of the $F$ functions, the expected deviation
by using in the right-hand side of eq.(\ref{ff3}) the crude
approximation $b_A\simeq b_B$, does not affect significantly this correction. 
In any case, the magnitude of the relevant correction,
$(F_B/F_A)^{1/2}$, is extremely small, typically: $\sim 0.8\%$ at $z=0.24$
dropping to $\sim 0.1\%$ at $z=2.1$, and the overall scaling of the
bias to different cosmologies is dominated by the ratio of the
corresponding $\sigma_8(z)$ variances.

We can facilitate our scaling procedure by using the $\sigma_8(z=0)$
power-spectrum normalizations of the different models, a value
always provided by the different authors. We therefore
translate the values of $\sigma_8(z)$ to
that at $z=0$ by using the linear growing mode of perturbations according to:
$\sigma_8(z)=\sigma_8(0) D(z)$. 
The final scaling relation from the $B$ cosmology to that of $A$, therefore becomes:
\be
b_A(z)\simeq b_B(z)\frac{\sigma_{8,m,B}(0)}{\sigma_{8,m,A}(0)} \frac{D_B(z)}{D_A(z)} 
\left[\frac{F(\Omega_{m,B},\Omega_{\Lambda,B},b_B,z)}{F(\Omega_{m,A},\Omega_{\Lambda,A},b_B,
  z)}\right]^{1/2} 
\;.
\ee
	
\section{Theoretical bias models}
Here we briefly describe the bias evolution models that we are going
to compare. As discussed in the introduction, we
separate the models in two families. The {\em galaxy merging model} 
family, based on the Press-Schether formalism and the
peak-background split. The models that we will investigate,
representing this family, are the Sheth, Mo \& Tormen (2001)
extension of the original Sheth \& Tormen (1998) model (hereafter
SMT), the Jing (1998) model, the Tinker et. al. (2010) (hereafter
TRK)
and the Ma et. al. (2011) model (hereafter MMRZ).
All these models provide the bias of halos as a function of the
peak-height parameter, $\nu$, where 
\be\label{eq:nu} 
\nu\equiv \delta_{c}(z)/\sigma(M_h,z)\;,
\ee
with $M_h$ the halo mass, $\sigma^2(M_h,z)$ the variance of the
mass fluctuation field at redshift $z$, and 
$\delta_c(z)$ the critical linear overdensity for spherical collapse,
which has a weak redshift dependence (see eq.18 of Weinberg \&
Kamionkowski 2003).

The basic free parameter of these bias models, to be fitted by the
data (although depending on the model one more parameter may be
allowed to vary - see below), is $\nu$ and through the evolution of
$\sigma(M_h,z)$ we will be able to derive the predicted bias redshift evolution, 
as well as the value of $M_h$. The latter value will be estimated by
using the definition of $\sigma$, eqs.(\ref{eq:spat1}) 
and (\ref{eq:PS}), from which we have that:
\be\label{eq:hmass}
\sigma^2(M_h)=\sigma^2_8 
\frac{\int^{\infty}_{0}dkk^{n+2}T^2(k)W^2(kR)}{\int^{\infty}_{0}dkk^{n+2}T^2(k)W^2(kR_8)}
\;,\ee
with $R=(3 M_h/4 \pi \bar{\rho})^{1/3}$, $R_8=8 \; h^{-1}$ Mpc and $\bar{\rho}=2.78
\times 10^{11} \Omega_m h^2 M_{\odot}/{\rm Mpc}^3$.

The second family contains the so-called {\em galaxy conserving} models and their
extensions. These models are based on the hydrodynamical equations of
motion and linear perturbation theory while the most general such model,
that we will investigate,
is that of Basilakos \& Plionis (2001; 2003), extended to included a
correction for halo merging in Basilakos, Plionis \& Ragone-Figuera
(2008).

Below we present the functional form of the bias evolution for
each of the models that we will investigate:

\subsection{BPR}
\indent

Basilakos and Plionis (2001; 2003) using linear perturbation theory and
the Friedmann-Lemaitre solutions derived a second-order
differential equation for the evolution of bias, assuming
that the mass-tracer population is conserved in time and that
the tracer and the underlying mass share the same dynamics.

The solution of their differential equation, 
for a flat cosmology, was found to be (Basilakos \& Plionis 2001):
\be
 b(z)=C_1E(z)+C_2E(z)I(z)+1
\ee
where $E(z)=\left[\Omega_m(1+z)^3+\Omega_{\Lambda}\right]^{1/2}$ and
\be
I(z)=\int^{\infty}_{z}\frac{(1+x)^3}{E^3(x)}dx \;.
\ee
The constants of integration depend on the halo mass, as shown in BPR, and
they are given by:
\be\label{eq:C1}
C_1(M_h)\approx \alpha_1\left(\frac{M_h}{10^{13}h^{-1}M_\odot}\right)^{\beta_1}
\ee
\be\label{eq:C2}
C_2(M_h)\approx \alpha_2 \left(\frac{M_h}{10^{13}h^{-1}M_\odot}\right)^{\beta_2}
\ee
and the values of $\alpha_{1,2}$ and $\beta_{1,2}$ where estimated
originally from a $\sim$WMAP1 $\Lambda$CDM numerical simulation in
BPR. We have since run a WMAP7 $\Lambda$CDM simulation, the
details of which can be found in Appendix A1, and from which we have
determined the new values of the $\alpha_{1,2}$ and $\beta_{1,2}$
parameters (see Table A1). 
The cosmological dependence of these parameters is also 
discussed in Appendix A2.

In BPR it was found that the original Basilakos \& Plionis model could
well reproduce the bias evolution for $z<3$, but not at higher
redshifts, indicating the necessity to extend the model to include the
contribution of an evolving mass-tracer population. 
Such an extension was presented in BPR and it was based on a
phenomenological approach, although the functional form for the effects
of merging was based on physically motivated arguments (see Appendix A2 of
BPR).
To this end they introduced to the continuity equation an additional
time-dependent term, $\Psi(t)$, associated with the effects of
merging of the mass tracers, which depends on the tracer  number
density, its logarithmic derivative and on $\delta_{tr}$. They
parameterized this term using a standard evolutionary form:
\be
\Psi(z)=AH_0(1+z)^{\mu}
\ee
where $\mu$ and $A$ are positive parameters which engulf the
(unknown) physics of galaxy merging. The bias evolution is now given by:
\be
b(z)=C_1E(z)+C_2E(z)I(z)+y_p(z)+1
\ee
where the additional halo-merging factor, $y_p(z)$, is given by:
\begin{eqnarray}
y_p(z)=E(z)\int^{z}_{0}\tau(x) I(x) dx -E(z)I(z)\int^{z}_{0}\tau(x) dx
\end{eqnarray}
with $\tau(z)= f(z)E^2(z)/(1+z)^3$ and
$f(z)=A(\mu-2) (1+z)^{\mu}E(z)/D(z)$. The values of both $A$ and
$\mu$ have been fitted using $\Lambda$CDM numerical simulations (see
BPR) and it was
found that $\mu\simeq 2.5-2.6$ independent of the halo
mass, while $A$ increases with decreasing halo mass, with $A\simeq
0.006$ and 0 for intermediate  (ie., $10^{13} \mincir M_h\mincir 10^{13.8}
h^{-1} M_{\odot}$) and higher mass halos, respectively. Evidently, the
bias factor at $z=0$ is provided by:
\be
b(z)=C_1+C_2 I(z)+1 \;.
\ee

Therefore in the current analysis we will leave $M_h$ as a free parameter to be
fitted by the data (BPR model) but we will also allow (a) the parameter $\alpha_1$ to
be fitted by the data, keeping $A$ equal to its simulation based value
($A=0.006$, BPR-I model), as well as the parameter $A$ to be fitted by the data
keeping $\alpha_1$ equal to its simulation based value
($\alpha_1=4.53$, BPR-II model). 

\subsection{SMT}
\indent

In Sheth et. al. (2001) the original work  of Sheth \& Tormen (1999)
was extended for the case of an ellipsoidal, rather than a spherical collapse. 
This new ingredient reduces the difference between theoretical expectations and
simulation DM halo data. Considering
ellipsoidal collapse the density threshold required for
collapse, contrary to the spherical collapse case, depends on the mass of
the final object.
		
Using the ratio of
the halo power spectrum to that of the underlying mass, they derived
the functional form for the bias as:
\begin{eqnarray}
b(\nu)=1+\frac{1}{\sqrt{a}\delta_{c}(z)} 
\left[\sqrt{a}(a\nu^2)+\sqrt{a}b(a\nu^2)^{1-c} -f(\nu)\right] \nonumber\\
{\rm with} \;\; f(\nu)=\frac{(a\nu^2)^c}{(a\nu^2)^c+b(1-c)(1-c/2)} \;,
\end{eqnarray}
where the free parameters where evaluated
using N-body simulations to have values: $a=0.707, b=0.5$ and $c=0.6$.
In particular the value of $a$ was found to
depend mostly on
how the simulation DM halos were identified. In the case of a Friends
of Friends (FoF) algorithm the value $a=0.707$ corresponds to the standard
linking length of 0.2 times the mean inter-particle
separation. Decreasing the linking length would increase the value of $a$ and
vice-versa (see discussion in SMT). Therefore, beyond the value of the
DM halo mass, $M_h$ (which will be estimated from the resulting value
of $\sigma(M_h)$ via eq.(\ref{eq:hmass}), we will also allow the
parameter $a$ to be fitted by the data. 

\subsection{JING}
\indent	
Jing (1998) used the clustering of simulation DM halos 
to derive an expression for the bias which is independent of the shape of
the initial power-spectrum, being CDM or power-law.
His corresponding expression is:
\be
b(\nu)=\left(\frac{0.5}{\nu^{4}}+1\right)^{(0.06-0.02n)}
\left(1+\frac{\nu^{2}-1}{\delta_{c}}\right)\;,
\ee	
where $n$ is the linear power spectrum index at the halo scale (ie., $n=d\ln
P(k)/d\ln k \simeq -2$ for $M_h\simeq 10^{13} h^{-1} M_{\odot}$). The
only free parameter of this model, to be fitted by the data, is the
halo mass, $M_h$ (which will be estimated from the fitted value of 
$\sigma(M_h)$ via eq.\ref{eq:hmass}).

\subsection{TRK}
\indent
Tinker et. al. (2010) measure the clustering of dark matter halos 
based on a large
series of collisionless N-body simulations of the $\Lambda$CDM
cosmology. DM halos were identified using the spherical
overdensity algorithm by which halos are considered as
isolated peaks in the density field such that the mean density is
$\Delta$ times the density of the background.
Their bias fitting function reads as:
\be
b(\nu)=1-A\frac{\nu^a}{\nu^a+\delta_c^a}+B\nu^b+C\nu^c \;
\ee
where $y=\log_{10}\Delta$.
For the WMAP7 $\Lambda$CDM model the value which corresponds
to the virialization limit is $\Delta_{\rm \rm \Lambda CDM}\simeq 355$. 
The rest of the parameters of the model are:
$A=1+0.24y \exp[-(4/y)^4]$, $B=0.183$,
$C=0.019+0.107y+0.19 \exp[-(4/y)^4]$, 
$a=0.44y-0.88$, $b=1.5$, $c=2.4$.

Therefore, we will fit the observational data using as a single free
parameter the DM halo mass ($M_h$, derived via $\sigma(M_h)$ in
eq.\ref{eq:hmass}) and using with $y=\log_{10}(\Delta_{\rm \Lambda
  CDM})$. However, we will also allow the latter parameter to be
fitted by the data, simultaneously with $M_h$.

\subsection{MMRZ}
\indent
	Ma et. al. (2011) extended the original Press-Schether approach
        incorporating a non-Markovian extension with a stochastic barrier,
        where they assume that the critical value for spherical collapse is
        itself a stochastic variable, whose scatter reflects a number
        of complicated aspects of the underlying dynamics. 
	Their model contains two parameters: $\kappa$, which
        parameterizes the degree of non-Markovianity and whose exact
        value depends on the shape of the filter function used to
        smooth the density field, and $\alpha$, the so-called {\em
          diffusion coefficient}, which parameterizes
        the degree of stochasticity of the barrier. Taking into
        account the non-Markovianity and the stochasticity of the
        barrier, the bias takes the form:
\begin{eqnarray}
b(\nu)=1+\frac{\alpha\nu^2-1+\frac{\alpha\kappa}{2}\left[2-e^{\alpha\nu^2/2}
\Gamma(0,\frac{\alpha\nu^2}{2})\right]}
{\sqrt{\alpha}\delta_c \left[1-\alpha\kappa+
\frac{\alpha\kappa}{2}e^{\alpha\nu^2/2}\Gamma(0,\frac{\alpha\nu^2}{2})\right]}
\end{eqnarray}
where $\alpha=(1+D_B)^{-1}$, with $D_B$ the diffusion coefficient, and 
$\Gamma(0,x)$ the incomplete gamma function.
Without the stochasticity of the barrier one has $D_B=0 \rightarrow
\alpha=1$.

Ma et
al. (2011) have found using N-body simulations that using
$\alpha=0.818$ and $\kappa=0.23$ they can reproduce to a good extent
both the simulation bias and the halo mass-function as a function of
$\nu$. We will therefore use these parameter values to fit the
observational bias data in order to constrain $M_h$. Additionally, we
will allow both $\alpha$ and $M_h$ to be fitted simultanesouly by the data, using
$\kappa=0.44$, since this is the value for a top-hat
smoothing kernel in coordinate space. Note that the value of $\kappa$
appears to be almost independent of cosmology, as discussed in
Maggiore \& Riotto (2010).

\section{Fitting Models to the Data}
\indent

In order quantify the free parameters of the DM halo bias models
we perform a standard $\chi^2$ minimization procedure between $N$ bias
data measurements, $b_i(z)$, with the bias values predicted 
by the models at the corresponding redshifts, $b({\bf p},z)$. 
The vector ${\bf p}$ represents the free parameters of the model 
and depending on the model their number is one or two. 
This procedure makes the simplistic
assumption that each DM halo hosts one mass tracer, an assumption
which is justified from the way the QSO and
galaxy bias data have been estimated (see discussion in section 2).

The $\chi^2$ function is defined as:
\be
\chi^2=\sum^{N}_{i=1}\left[\frac{b_i(z)-b({\bf p},z)}{\sigma_{b_{i}}(z)}\right]^2\;,
\ee
with $\sigma_{b_{i}}(z)$ is the observed bias uncertainty. 
We have in total $N=22$ measured bias data for the optical QSOs,
spanning from $z=0.24$ to $z=4$, and
$N=5$ for the optical galaxies, spanning from $z=0.55$ to $z=1.4$.

Note that the uncertainty
of the fitted parameters will be estimated, in the case of more than
one such parameter, by marginalizing one with respect to the other. 
However, since such a procedure may hide possible
degeneracies between parameters, we will also present the 1, 2 and
3$\sigma$ likelihood contours in the parameter plane.

Furthermore, since we will attempt to compare the different models among them, the
$\chi^2$ test alone is not sufficient for such a task, 
since different models may have a different number of free
parameters. Instead we will use information criteria to compare the
strengths of the different models, according to the
work of Liddle (2004), a procedure that favors those models that give
a similarly good fit to the data but with fewer free parameters (see
for example Saini et al. 2004; Godlowski \& Szydlowski 2005; Davis et
al. 2007 and references therein). To this end we will use, the relevant to our case, 
{\em corrected} Akaike information criterion for small sample size 
(${\rm AIC}_c$; Akaike 1974, Sugiura 1978), defined,
for the case of Gaussian errors, as:
\be
{\rm AIC}_c=\chi^2+2k+2k(k-1)/(N-k-1)
\ee
where $k$ is the number of free parameters, and thus when $k=1$ then
AIC$_c=\chi_{\rm min}^2+2$. A smaller value of AIC$_c$
indicates a better model-data fit. However, small
 differences in AIC$_c$ are not necessarily significant and therefore, in order
 to assess, the effectiveness of the different models in reproducing
 the data, one has to investigate the model pair difference 
$\Delta$AIC$_c = {\rm AIC}_{c,y} - {\rm AIC}_{c,x}$. 
The higher the value of $|\Delta{\rm AIC}_c|$, the
higher the evidence against the model with higher value of ${\rm AIC}_c$,
with a difference $|\Delta$AIC$_c| \magcir 2$ indicating a positive such evidence and
$|\Delta$AIC$_c| \magcir 6$ indicating a strong such evidence,
 while a value $\mincir 2$ indicates consistency among the two 
comparison models.

\subsection{Optical QSO Results}
Here we fit the different bias evolution models to the
scaled to the WMAP7 cosmology optical QSO data, described in
section 2. 
It is important to note that all the bias models used in this work
(except the BPR) have been studied as a function of the threshold
$\nu$, eq.(\ref{eq:nu}), ie., in effect as a function of the variance of the fluctuation
field and thus as a function of halo mass, while the free parameters of most models have been 
fitted using $z=0$ simulations. In these models the redshift dependence of the bias
comes mostly from the redshift dependence of the peak-height, $\nu$
(see eq.\ref{eq:nu}).

We will present separately the results of the models with
one free parameter, the halo mass, and the models with an additional
free parameter, as discussed in the theoretical model presentation sections.
\begin{table}
\tabcolsep 8pt
\begin{tabular}{llccc}  \hline
Model &$10^{12} h^{-1} M_{\odot}$ &$b(0)$&$\chi^2_{\rm min}/df$&AIC$_c$  \\ \hline
BPR    &$3.0\pm 0.4$                   &1.02&      $12.88/21$ & $14.88$  \\
SMT    &$3.2\pm 0.4$                   &1.07&      $23.21/21$ & $25.21$  \\
JING   &$2.1\pm 0.2$                   &0.98&      $18.00/21$ & $20.00$  \\ 
TRK    &$3.0\pm 0.4$                   &1.00&      $15.79/21$ & $17.79$   \\ 
MMRZ   &$2.2\pm 0.2$                   &0.87&      $20.14/21$ & $22.14$   \\ \hline
\end{tabular}	
\caption{Results of the $\chi^2$ minimization procedure between the
  optical QSO data ($N=22$) and bias models with one free
  parameter ($k=1$).}
\end{table}

\begin{figure}
\resizebox{8.5cm}{8.5cm}{\includegraphics{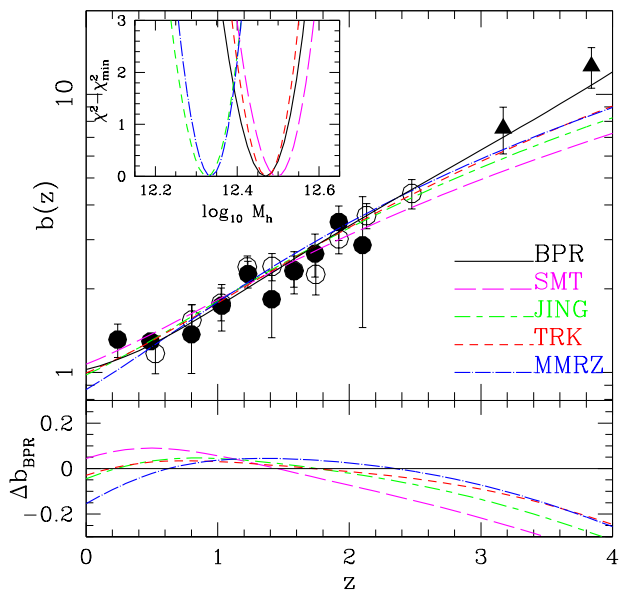}}
\caption{{\em Main Panel:} Comparison of the QSO bias data 
(open circles correspond to Croom et. al. 2005; filled circles to
Ross et al. 2009 and filled triangles to the high redshift data of 
Shen et. al. 2010), with the one free parameter bias model fits (BPR: black
continuous line; SMT: magenta dashed
line; JING: green long-short dashed line; TRK: red short-dashed line;
MMRZ: blue dot-dashed line). {\em Lower Panel:}
The relative difference between the BPR model and all the rest, 
$\Delta b_{\rm BPR}(z)$.
{\em Inset Panel:} The value of $\chi^2-\chi^2_{\rm min}$ as a function
of halo mass, $M_h$, for the indicated bias models.}
\end{figure}

\subsubsection{One free parameter models}
In Table 1 we present the best fit model parameters based on the
$\chi^2$ minimization procedure, with the first and second columns
listing the fitted halo mass, $M_h$,
derived using eq.(\ref{eq:hmass}) and the value of bias at $z=0$,
respectively. We also present the goodness of fit statistics, as
discussed previously (reduced $\chi^2$ and AIC$_c$).
In the main panel of Figure $1$ we present the bias evolution models (different
 lines), using the best fit parameter values listed in Table 1
together with the WMAP7-scaled optical QSO bias data. The inset panel
of Fig.(1) shows that the resulting $M_h$ values cluster around two,
relatively similar, values: $\sim 3 \times 10^{12} h^{-1} M_{\odot}$ and $\sim
2 \times 10^{12} h^{-1} M_{\odot}$. In the lower panel we present
the relative difference between the BPR model and all the rest, 
ie., $\Delta b_{\rm BPR}(z)=[b_{i}(z)-b_{\rm BPR}(z)]/b_{\rm BPR}(z)$.

Some basic conclusions that become evident, inspecting also Table 1, are:
\begin{itemize}
\item Although all one free parameter bias models appear to fit at a
  statistically acceptable
  level the optical QSO bias data, by far the best model is the BPR,
  which is the only model fitting also the highest redshifts ($z\magcir
  3$). The MMRZ is the only model that does not fit the lowest
  redshifts ($z\mincir 0.3$), providing an anti-biased value at the
  current epoch, $b(0)=0.87$.
\item The relative bias difference of the various fitted models with
  respect to that of BPR, $\Delta b_{\rm BPR}(z)$, 
  indicates that the BPR, JING and TRK models have a very similar
  redshift dependence for $z \mincir 2.2$ (with $|\Delta b_{\rm BPR}(z)| \mincir
  0.05$), while all the models show very large such deviations
  for $z\magcir 2.8$, reaching $|\Delta b_{\rm BPR}|\magcir 0.3$ at
  the largest redshifts. The SMT and MMRZ models show large deviations
  at the lowest redshifts as well.
\item Beyond the fact that the BPR model provides by far the best fit to the
  QSO bias data, the second best model is the TRK model, with
  $\Delta$AIC$_c\sim -2.9$. Furthermore,
  one can distinguish that the model pairs (JING, TRK) and (JING,
  MMRZ) are statistically equivalent ($\Delta$AIC$_c\mincir 2$).
\item The traditional SMT and the recently proposed MMRZ models rate the worst
  among all the other one parameter models, but interestingly the
  former provides consistent values of $M_h$ and $b(0)$ with those of
  the BPR model.  
\end{itemize}

We attempt now to provide a robust average value of the DM
halo mass that hosts optical QSO, using an inverse-AIC$_c$ weighting
of the different one parameter model results. This procedure provides a
weighted mean and combined weighted standard deviation of the DM halo mass of:
$$(\mu_{M_h}, \sigma_{M_{h}}) = (2.72, 0.56) \times 10^{12} h^{-1} M_{\odot}$$
while the weighted scatter of the mean is $\sim 0.44\times 10^{12} h^{-1} M_{\odot}$.

Finally, we point out that since it appears that mainly the 2
high-$z$ bias points are the ones that give the advantage to the BPR
model with respect to the others, we perform 
a more conservative comparison among the models
by excluding these two high-$z$ data points.
We find that although the resulting halo mass and $b(0)$ 
are very similar to those of Table 1, with variations of a few percent, 
there are now three models that perform equivalently well, the BPR,
JING and TRK with AIC$_c\simeq 13$. The other two models perform
moderately (SMT) or significantly (MMRZ) worse, as was the case also
in the full data comparison, with $\Delta$AIC$_c \simeq 2$ and 4, respectively.

\subsubsection{Two free parameter models}
We now allow a second parameter to be fitted simultaneously with the
DM halo mass. Since the free parameters of the bias models have been
determined using N-body simulations, it would be interesting to
investigate if their simulation-based value can be reproduced by real
observational data. 
The second free parameter that we will use is $\alpha_1$, $A$, $a$, $y$ and  $\alpha$ for
the BPR-I, BPR-II, SMT-I, TRK-I and MMRZ-I models, respectively. Note
that in the case of the BPR-II model we will use the simulation based
value of $\alpha_1$, with the free parameter $A$ being the
halo-merging parameter of the BPR model (defined in section 3.5).
\begin{table}
\tabcolsep 3.5pt
\begin{tabular}{llcccc}  \hline

Model &$10^{12} h^{-1} M_{\odot}$ &2$^{\rm nd}$param.&$b(0)$&$\chi^2_{\rm min}/df$&AIC$_c$ \\ \hline
BPR-I & $2.2\pm 0.4$    &$4.64\pm0.07$   &1.08 &        $11.95/20$&$16.16$  \\
BPR-II& $2.8\pm 0.6$    &$0.008\pm 0.005$&1.02 &        $12.45/20$&$16.66$  \\
SMT-I & $27.0\pm 4.0$   &$0.40\pm0.02$   &0.95 &        $18.72/20$&$22.93$  \\
TRK-I & $0.4\pm 0.1$    &$6.14\pm 0.43$  &1.11 &        $13.37/20$&$17.58$  \\  
MMRZ-I& $0.3\pm 0.1$    &$1.21\pm 0.04$  &0.87 &        $21.27/20$&$25.48$  \\ \hline
\end{tabular}	
\caption{Results of the $\chi^2$ minimization procedure between the
  optical QSO data and the bias models with 2 free parameters.}
\end{table}

Table 2 presents the best fit model parameters resulting from the
$\chi^2$ minimization procedure, with the first and second columns
representing respectively the resulting halo mass, $M_h$,
and the second free parameter, while the third column the value of the
bias at $z=0$.
In Figure 2 we compare the resulting bias evolution models with the
WMAP7 scaled QSO bias data (as in Figure 1), while in the lower panel
we present the relative difference between the BPR-II model and all the rest, 
ie., $\Delta b_{\rm BPR-II}(z)=[b_{i}(z)-b_{\rm BPR-II}(z)]/b_{\rm BPR-II}(z)$.

Below we list the main conclusions of the above fitting procedure:
\begin{itemize}
\item A first important result is that the only model that reproduces the
simulation-based second free parameter value, is the BPR-I model. The
simulation based value is $\alpha_1=4.53$ while the fitted value, based
on the QSO bias data, is $\alpha_1=4.64\pm 0.07$. 
This fact will allow us to derive
the dependence of the parameters of the BPR bias evolution model 
on the relevant cosmological parameters
(see Appendix A2). 
\item Fitting the BPR-II model to the QSO data provides $A=0.008\pm
  0.005$ which is almost identical to the simulation
  determined value, used in the BPR case ($A=0.006$). As it is
  therefore expected, the fitted values of $M_h$ and $b(0)$ are almost identical
  to those of the one parameter BPR model, but the statistical
  significance of the BPR-II model is lower than that of the BPR due
  to its 2 free parameters.
\item Although the SMT-I and TRK-I models appear now to
fit slightly better the QSO bias data,
especially the higher $z$ range, this happens on the expense of providing
unexpected values for the $M_h$ and very different values of the second
fitted parameter with respect to their simulation based value. For example,
the SMT-I model provides a huge halo mass, $\sim$9 times larger than that
provided by the corresponding one parameter model.
This should be attributed to the fact that 
the fitted second parameter, $a$, is significantly
smaller than the nominal value of 0.707.
Similarly, the TRK-I model provides a very small value of
$M_h$, a factor of $\sim 9$ less than of the corresponding one parameter
model, while the resulting value $y=6.14$ implies that
$\Delta\simeq 10^{6}$, a value extremely large and unphysical. 
Finally, the MMRZ-I model provides again a very small value of $M_h$,
while it is the only two-parameter model that fits
the data worst than the corresponding one parameter model. This is due to the fact
that we have used $\kappa=0.44$ and not $\kappa=0.23$, which is used in the one
free parameter model, as suggested by MMRZ. Had we used the
latter $\kappa$ value we would have found an extremely small value of
$M_h\simeq 10^{10} h^{-1} M_{\odot}$.
These results probably indicate a degeneracy between the two fitted
parameters, a fact which we indeed confirm for the SMT-I, TRK-I and MMRZ-I
models, as can be seen in Figure 3 where we plot the 1,
2 and 3$\sigma$ likelihood contours in the parameter solution
plane. Contrary to the above models, no such degeneracy is present for
BPR-I model. 
Note that in Fig.3 the cross indicates the best 2-parameter solution, while
the dashed line indicates the simulation based value of the second parameter.
In the case of the TRK-I model the latter corresponds to the virialization
value ($y=\log_{10} 355$), used in the one free parameter fit.
\begin{figure}
\resizebox{8.5cm}{8.5cm}{\includegraphics{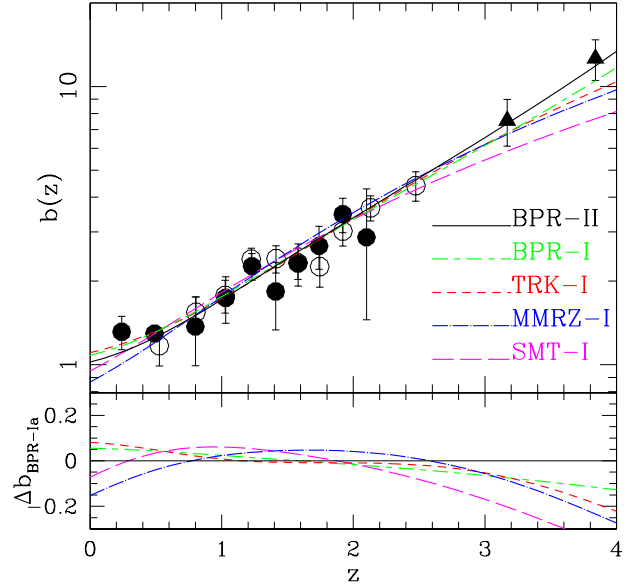}}
\caption{Comparison of the QSO bias data with the two free parameter
  bias models (their line types and colors are indicated in the Figure).
{\em Lower Panel:} The relative difference, $\Delta b_{\rm BPR-II}(z)$, 
between the BPR-II and the rest of the models.}
\end{figure}
\item Beyond the previously mentioned fundamental flow of the two
  parameter models (SMT-I, TRK-I and MMRZ-I), they all provide 
  relatively comparable to the BPR-II model fits of the QSO bias data
  but only within the range
  $0.8\mincir z\mincir 2.4$ (see lower panel of Fig.2). Furthermore, the
  MMRZ-I model, as in the case of the one free parameter fit, 
  provides an anti-biased value at $z\mincir 0.2$, while it
  also provides the worst overall fit to the QSO bias data.
\item Finally, the BPR one parameter model scores the best among all
  the one or two free parameter models and over the whole available
  QSO bias redshift range, while it is statistically equivalent with 
  the BPR-I and BPR-II models (since $|\Delta$AIC$_c|\mincir 1.8$; see
  Table 3).
\end{itemize}

\begin{figure}
\resizebox{4.cm}{4.cm}{\includegraphics{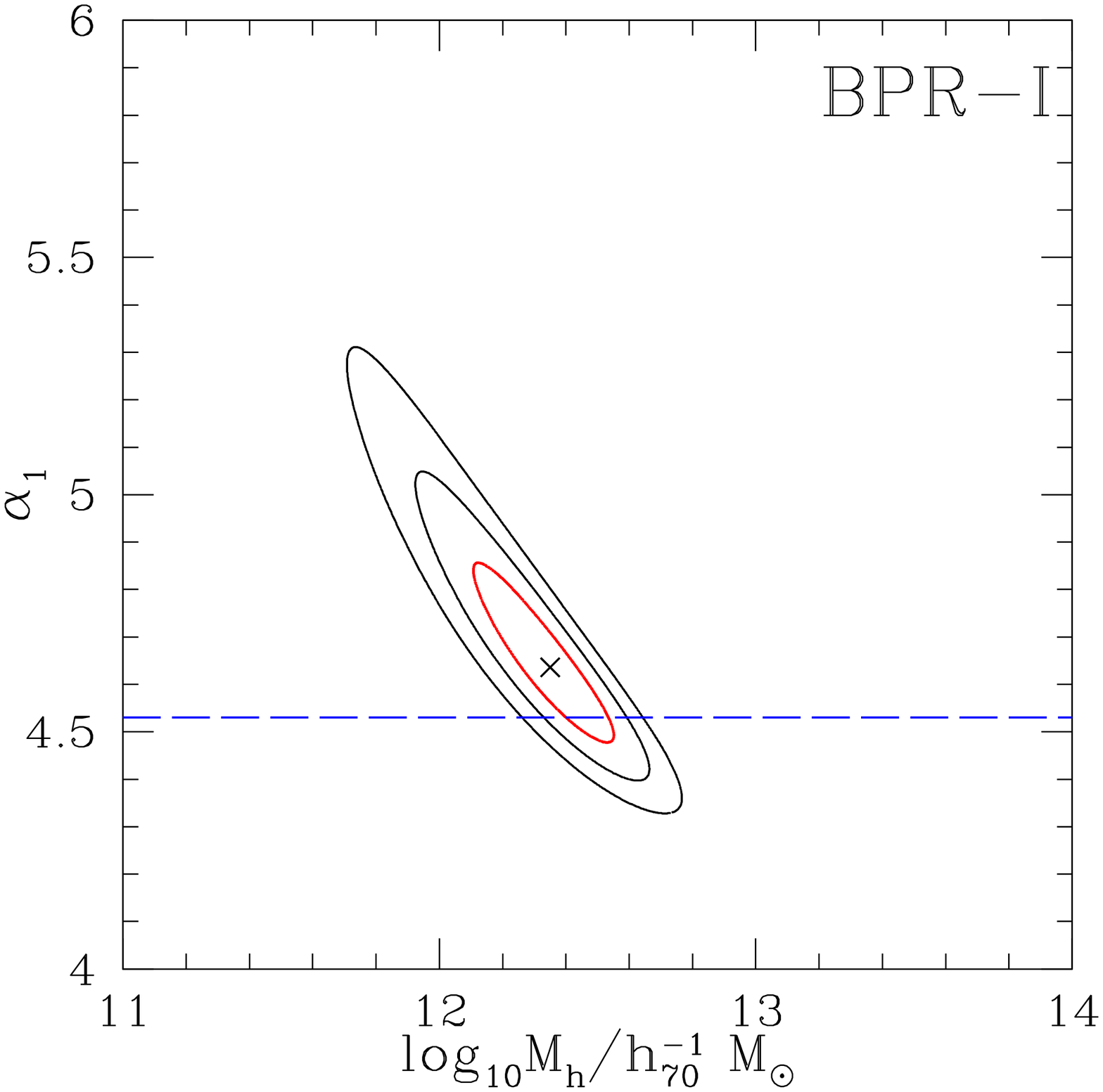}} \hfill
\resizebox{4.cm}{4.cm}{\includegraphics{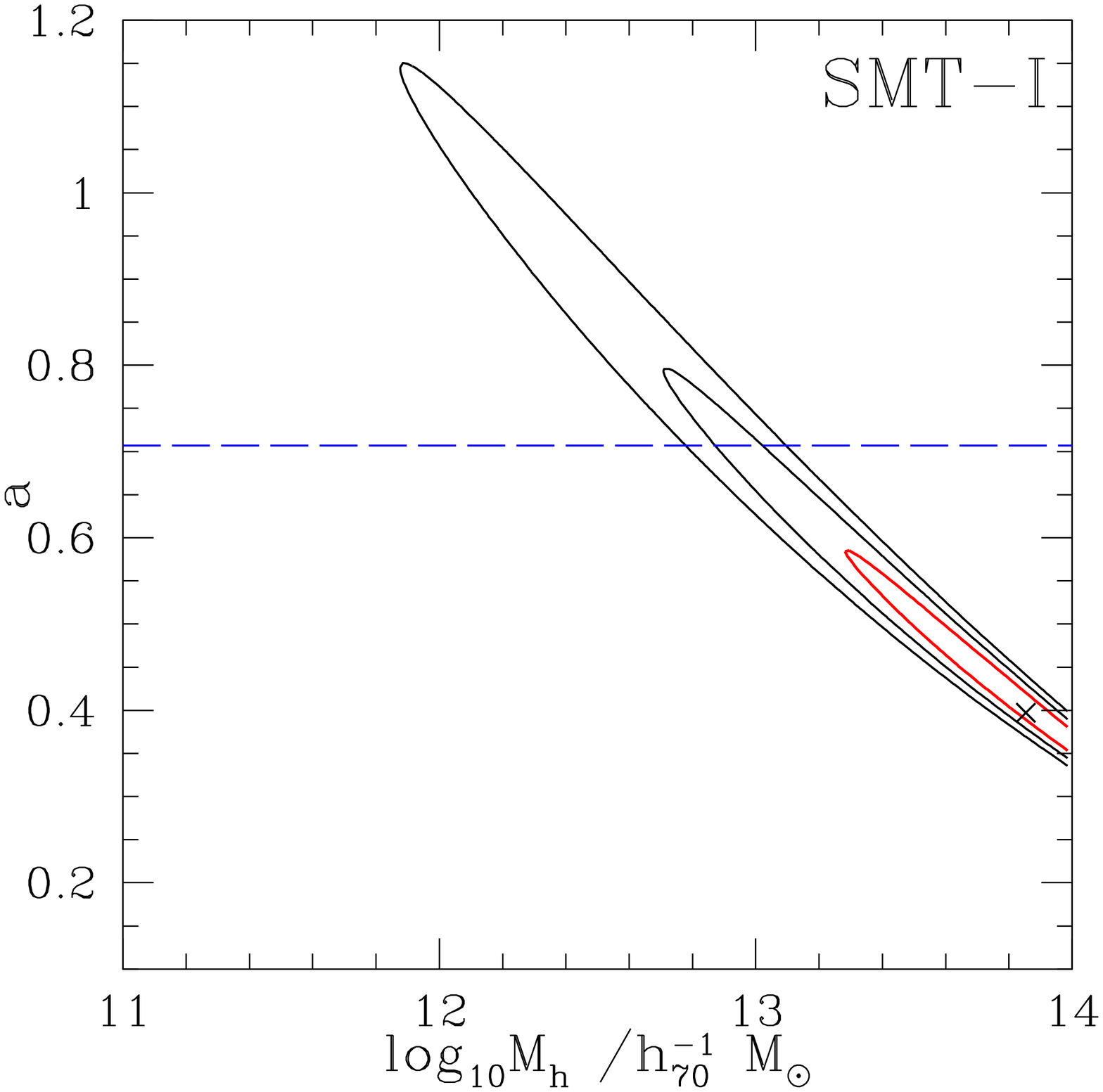}} \vfill
\resizebox{4.cm}{4.cm}{\includegraphics{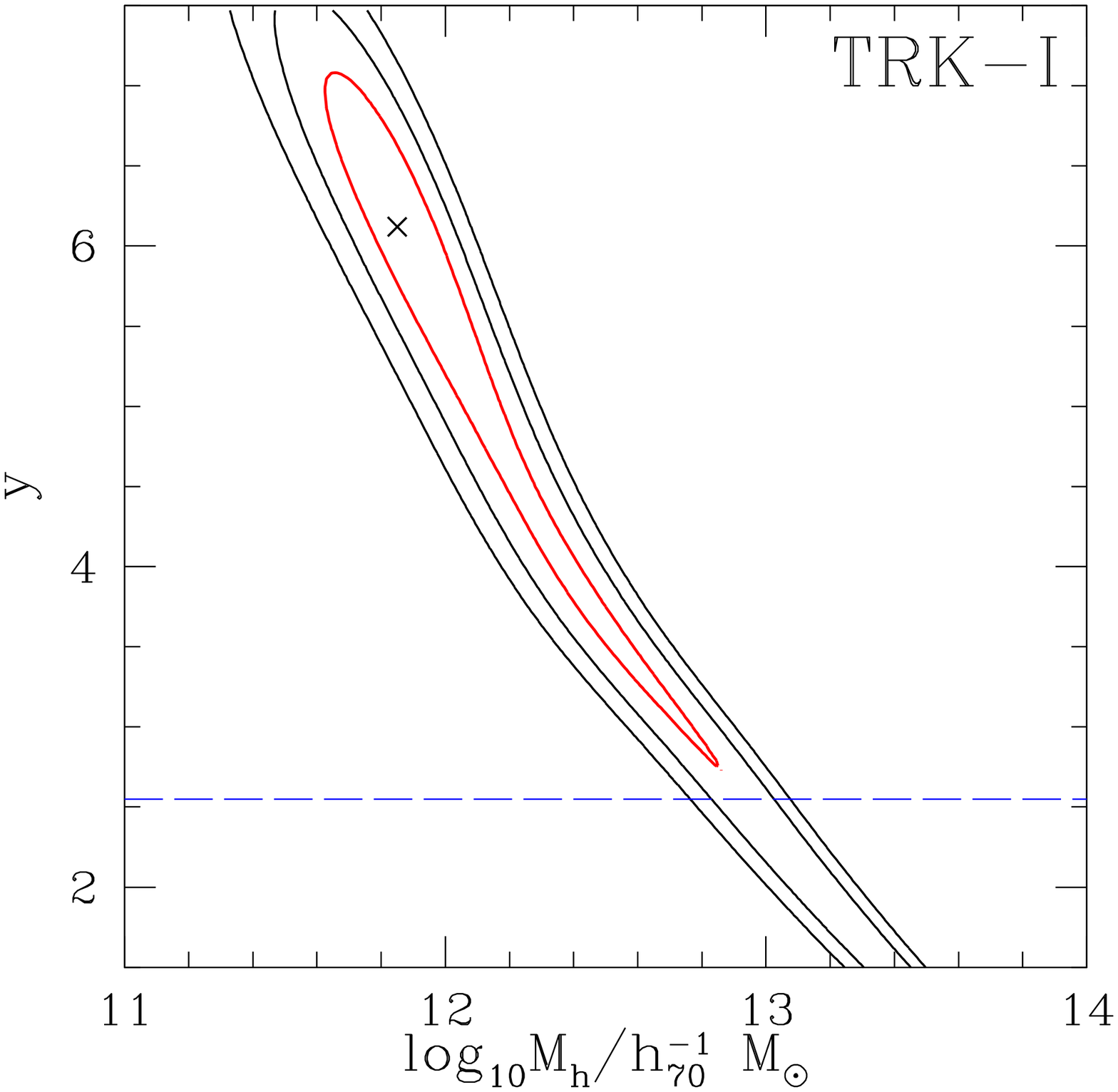}} \hfill
\resizebox{4.cm}{4.cm}{\includegraphics{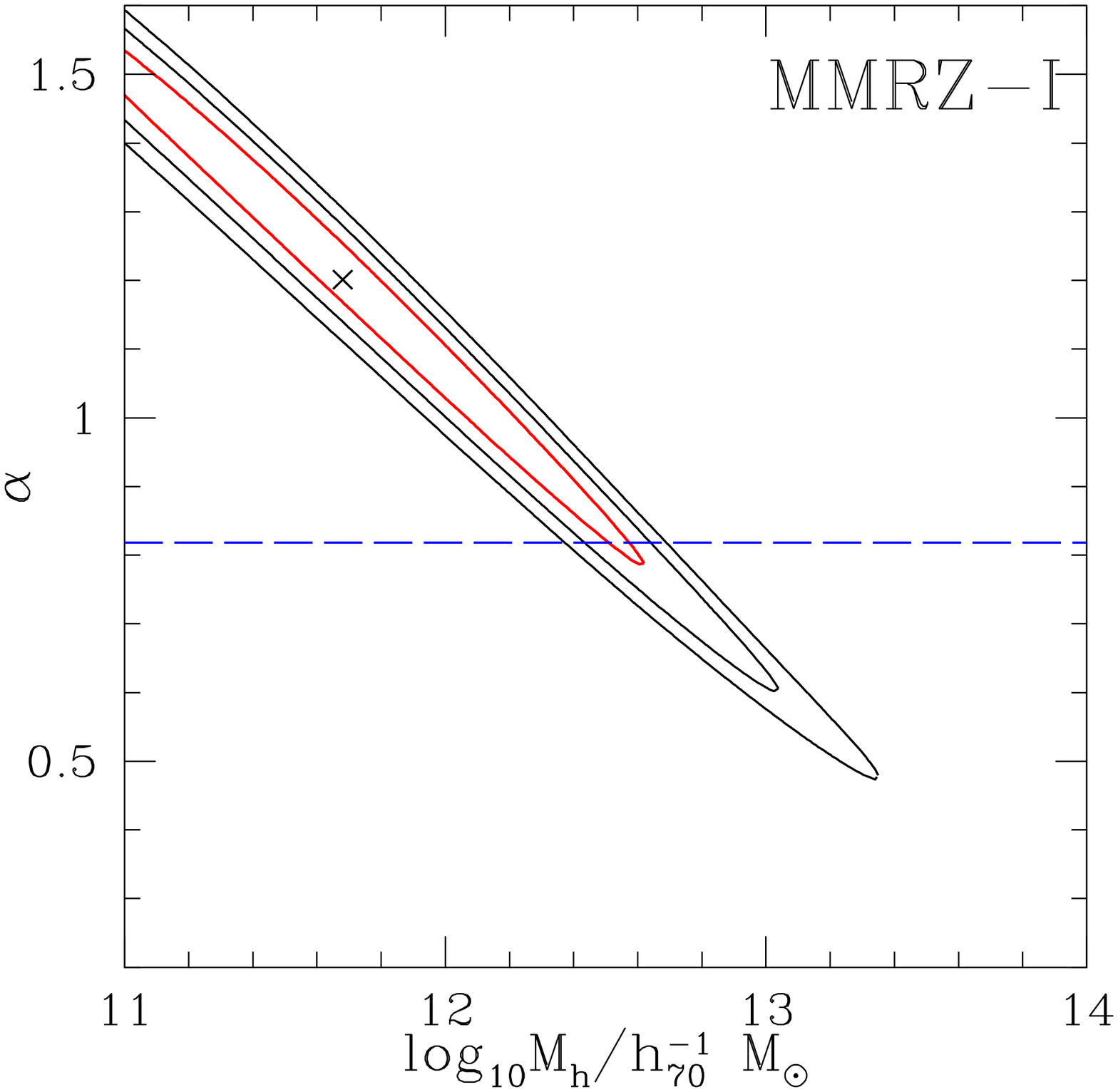}} 

\caption{Contour plots of the two fitted parameter solution
  space. The 1$\sigma$ level is indicated by the relatively thicker
  red curve. The blue dashed line indicates the simulation based value of
  the y-axis parameter.}
\end{figure}

We assess in a more quantitative manner the statistical relevance of the
different theoretical bias models in representing the observational
QSO bias data by using the information theory parameter
AIC$_c$ and presenting in Table 3 the model pair difference $\Delta$AIC$_c$. 
As previously discussed a smaller AIC$_c$ value indicates a model that
better fits the data, while a small $|\Delta$AIC$_c|$ value (ie., $\mincir 2)$
indicates that the two comparison models represent the data at a
statistically equivalent level.
\begin{table}
\tabcolsep 5pt
\begin{tabular}{lrrrrrr} \hline

        & BPR-I&BPR-II& SMT  & JING & TRK& MMRZ \\ \hline 
BPR     & -1.3 & -1.8 &-10.3 &-5.1 &-2.9 &-7.3\\ 
BPR-I   &      & -0.5 & -9.0 &-3.8 &-1.6 &-6.0\\    
BPR-II  &      &      & -8.5 &-3.3 &-1.1 &-5.5\\    
SMT     &      &      &      & 5.2 & 7.4 & 3.1\\    
JING    &      &      &      &     & 2.2 &-2.1\\      
TRK     &      &      &      &     &     &-4.3\\ \hline       
\end{tabular}	
\caption{Results of the pair difference $\Delta$AIC$_c$ for the bias
  evolution models fitted to the optical QSO data. }
\end{table}
Due to the resulting unphysical second free parameter, as discussed
previously, we do not use in Table 3 a comparison based on the
SMT-I, TRK-I and MMRZ-I models.
It is obvious that the one free parameter BPR model fairs the best
among any model, while it is statistically equivalent, as indicated by
the relevant values of $\Delta$AIC$_c$, to the BPR-I and BPR-II models
and to a slightly lesser degree to the TRK model. 

\subsection{Optical VVTS galaxy Results}
\indent
This model-data comparison takes place at relatively low redshifts
($z<1.5$), covering a small dynamical range in $z$, 
and therefore we will use only the one free parameter models
to fit the galaxy bias data. 
An additional reason is that even with the much larger
$z$-dynamical range covered by the QSO data, the second parameter could
not be constrained (except for the case of the BPR-I and BPR-II models).

The results of the $\chi^2$-minimization procedure are presented in Table
4, while in Figure 4 we present the model fits to the galaxy bias
data. Note that the layout of Figure 4 is as Figure 1.
\begin{table}
\tabcolsep 9pt
\begin{tabular}{llccc} \hline
Model &$10^{11} h^{-1} M_{\odot}$ &$b(0)$&$\chi^2_{min}/df$&AIC$_c$  \\  \hline
BPR   &$6.0\pm 2.5$ &0.99  &             $0.29/4$&2.29  \\
SMT   &$6.4\pm 1.9$ &0.90  &             $0.45/4$&2.45  \\
JING  &$5.1\pm 1.3$ &0.83  &             $0.93/4$&2.93  \\ 
TRK   &$7.8\pm 1.9$ &0.86  &             $0.74/4$&2.74 \\
MMRZ  &$6.1\pm 1.3$ &0.70  &             $2.10/4$&4.10   \\ \hline
\end{tabular}	
\caption{Results of the $\chi^2$ minimization procedure between the
  one free parameter models and  the optical VVTS galaxy bias data.}
\end{table}
It is evident that the BPR model fairs the best providing the
lowest reduced $\chi^2$ and AIC$_c$ parameter with respect to the
other models, while the MMRZ model fairs the worst.
However, due to the small dynamical range in redshift, the 
information theory pair model characterization parameter, $\Delta$AIC$_c$,
indicates that all the bias models are statistically equivalent in
representing the bias data, since $\Delta$AIC$_c\mincir 1.8$ for any
model pair. As in the QSO case, we provide an average halo mass that
hosts VVTS optical galaxies using an AIC$_c$ weighted procedure over
the different one-parameter bias models. The resulting weighted mean
and combined weighted standard deviation are:
$$(\mu_{M_h}, \sigma_{M_{h}}) = (6.3, 2.1) \times 10^{11} h^{-1} M_{\odot}$$
while the weighted scatter of the mean is also $\sim 0.9 \times 10^{11} h^{-1} M_{\odot}$.

It is interesting to point out that the only model that finds that at
$z=0$ the optical galaxies are unbiased ($b_0\simeq 1$), in agreement
with other studies of wide-area optical galaxy catalogues (Verde et
al. 2002; Lahav et al. 2002), is the BPR
model, while all the other models indicate that optical galaxies are quite
anti-biased with $b(0)\le0.9$.

\begin{figure}
\resizebox{8.5cm}{8.5cm}{\includegraphics{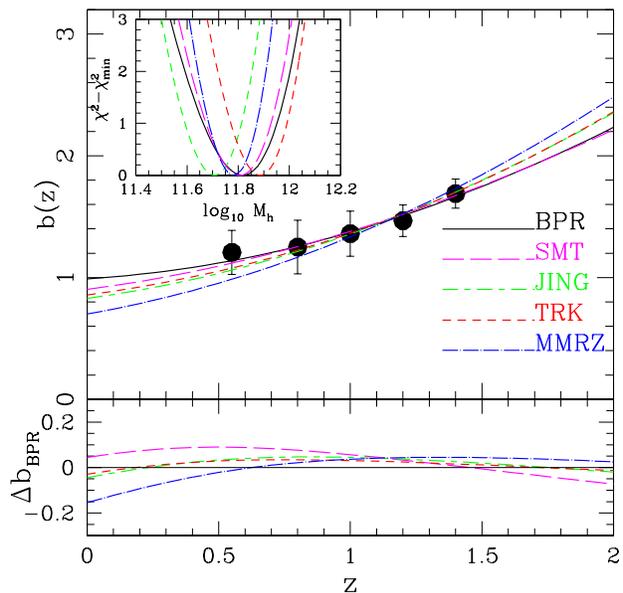}}
\caption{Results of the $\chi^2$ minimization procedure between the
  VVTS galaxy bias data and the bias models with one free
  parameter.}
\end{figure}

\section{Conclusions}
\indent
In this work we assess the ability of five recent bias
evolution models to represent a variety of observational bias data, based
either on optical QSO or optical galaxies.
To this end we applied a $\chi^2$ minimization procedure between
the observational bias data, after rescaling them to the WMAP7 cosmology, 
with the model expectations, through which we fit the model
free parameters.

In performing this comparison we assume that each halo is populated by
one extragalactic mass tracer, being a QSO or a galaxy; an assumption
which is justified since the observational data have been estimated on
the basis of either the large-scale clustering ($>1 h^{-1}$ Mpc),
corresponding to the halo-halo term, or the corrected for non-linear
effects variance of the smoothed (on 8 $h^{-1}$ Mpc scales) density field.

The comparison shows that all models fit at an acceptable level the
QSO data as indicated by the their reduced $\chi^2$ values.
Using the information theory characteristic, AIC$_c$, which takes into
account the different number of model free parameters we find that the
model that  rates the best among all the other is the Basilakos \&
Plionis (2001; 2003) model  with the tracer merging
extension of Basilakos, Plionis \& Ragone-Figueroa (2008),
which is the only model fitting accurately the optical QSO bias data
over the whole redshift range traced ($0<z<4$). The only other model
that is statistically equivalent at an acceptable level is that of Tinker et al. (2010).
The average, over the different bias models, DM halo mass that hosts optical QSOs is:
$M_h\simeq 2.7 (\pm 0.6) \times 10^{12} h^{-1} M_{\odot}$.

Finally, all the investigated bias models fit well and at a
statistically equivalent level the
VVTS galaxy bias data, with the BPR model scoring again the best,
and the MMRZ the worst.
The average, over the different bias models, DM halo mass hosting optical galaxies is:
$M_h\simeq 6 (\pm 2) \times 10^{11} h^{-1} M_{\odot}$.

\section*{Acknowledgements} 
S.B. wishes to thank the Dept. ECM of the
University of Barcelona for hospitality, and acknowledges financial support from the
Spanish Ministry of Education, within the program of Estancias de
Profesores e Investigadores Extranjeros en Centros Espa\~noles (SAB2010-0118).

\appendix
\section{Simulation based BPR model Parameter estimation}

\subsection{$\Lambda$CDM Simulations}
We have run a new WMAP7 $\Lambda$CDM N-body simulation using the
GADGET-2 code (Springel 2005) with dark matter only. 
The size of the box simulation is $500 h^{-1}$ Mpc and the number of
particles is $512^3$. The
adopted cosmological parameters are the following: $\Omega_m=0.273$
$\Omega_{\Lambda}=0.727$, $h=0.704$, $\sigma_8=0.81$ and the
particle mass is $7.07 \times 10^{10} h^{-1} M_{\odot}$, comparable to
the mass of a single galaxy. The
initial conditions were generated using the GRAFIC2 package
(Bertschinger 2001).
We also use a similar size simulation, generated in Ragone-Figueroa \& Plionis (2007),
of a $\Lambda$CDM model with $\Omega_{\rm m}=0.3$, $\Omega_{\Lambda} = 0.7$, $h=0.72$ and 
$\sigma_8=0.9$.

The dark matter haloes were defined using a FoF algorithm with a linking
length $l=0.17\langle n\rangle^{-1/3}$, where $\langle n \rangle$ 
is the mean particle density. 

We estimate the bias redshift evolution of the different DM haloes, 
with respect to the underlying matter distribution,
by measuring their relative fluctuations in spheres of radius 
8 $h^{-1}$ Mpc, according to the definition of eq.(\ref{bias2}), ie.,
\be
b(M,z)=\frac{\sigma_{8, h}(M,z)}{\sigma_{8, m}(z)} \;,
\ee 
where the subscripts $h$ and $m$ denote haloes and 
the underlying mass, respectively.
The values of $\sigma_{8,h}(M,z)$, for haloes of mass $M$, are computed
at different redshifts, $z$, by:
\be
\sigma^2_{8,h}(M,z)=\left\langle \left(\frac{N-\bar{N}}{\bar{N}} 
\right)^2\right\rangle-\frac{1}{\bar{N}} \;,
\ee 
where $\bar{N}$ is the mean number of such haloes in spheres of 8
$h^{-1}$ Mpc radius and the factor $1/\bar{N}$ is the expected
Poissonian contribution to the value of $\sigma^2_{8,h}$.
Similarly, we estimate at each redshift the value of the underlying
mass $\sigma_{8,m}$.
In order to numerically estimate $\sigma^2_{8,j}$ we randomly place 
$N_{\rm rand}$ sphere centers in the simulation volume, such that the sum of 
their volumes is equal to $\sim 1/8$ the simulation volume ($N_{\rm rand}=
8000$).  This is to ensure that we are not oversampling the available 
volume, in which case we would have been multiply sampling the same
halo or mass fluctuations. The relevant uncertainties are estimated
as the dispersion of $\sigma^2_{8,j}$ over 
20 bootstrap re-samplings of the corresponding halo sample. 
\begin{figure}
\resizebox{8.5cm}{4.5cm}{\includegraphics{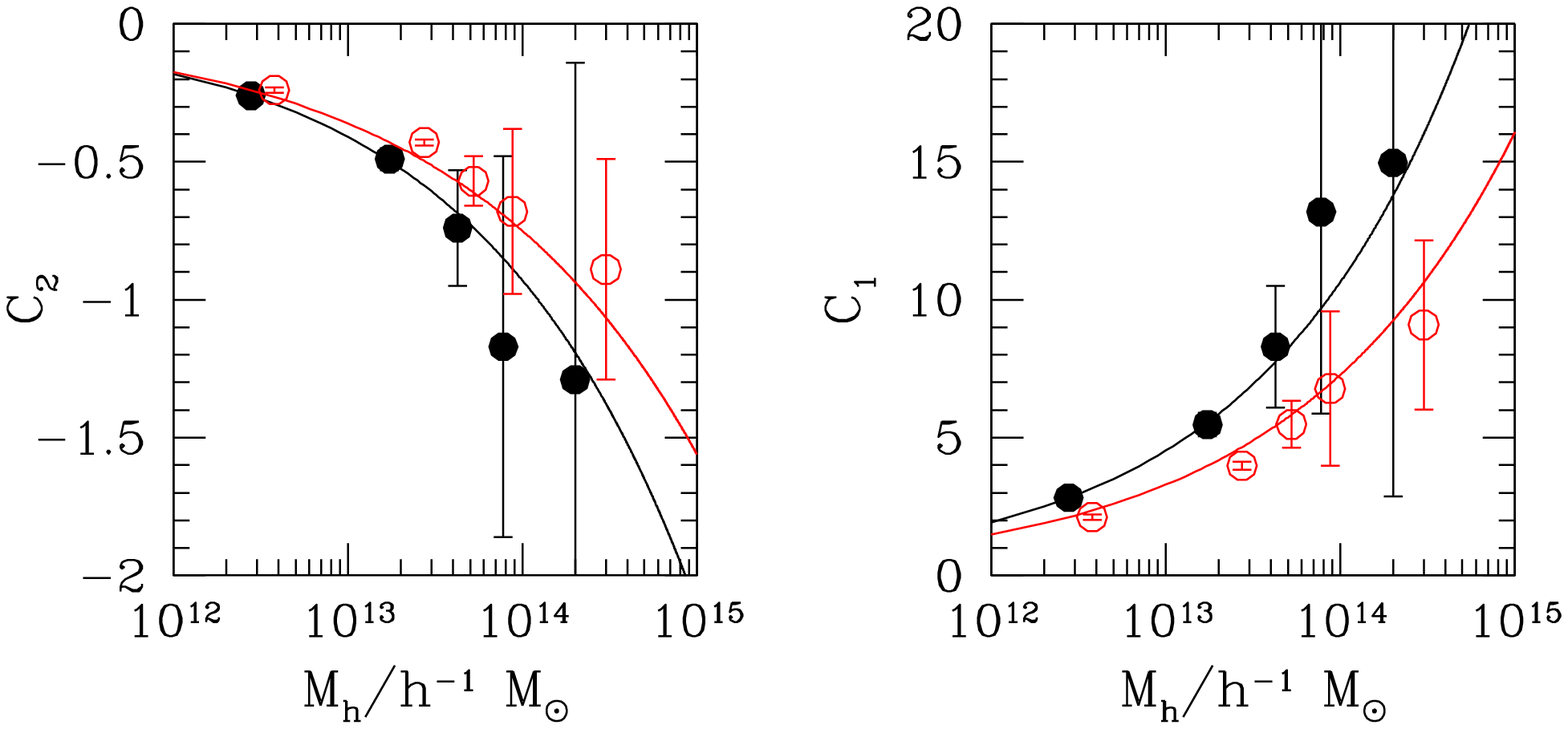}}
\caption{The Parameters ${\cal C}_1$ and ${\cal C}_2$ derived from the 
WMAP7 (filled points) and WMAP1 (open points) $\Lambda$CDM
  simulation (points). Continuous lines correspond to the function
  form of eq. (A3), with best fit parameters shown in Table A1.} 
\end{figure}
\begin{table}
\tabcolsep 5pt
\begin{tabular}{lcccc} \hline
Model &$\alpha_{1}$&$\alpha_{2}$&$\beta_{1}$&$\beta_{2}$ \\ \hline
WMAP1 & 3.30$\pm 0.13$ & -0.36$\pm 0.01$ & 0.34$\pm 0.04$ &0.32$\pm0.04$ \\
WMAP7 & 4.53$\pm 0.22$ & -0.41$\pm 0.02$ & 0.37$\pm 0.04$ &0.36$\pm0.04$ \\ \hline
\end{tabular}	
\caption{Results of the $\chi^2$ minimization used to evaluate the
  parameters that enter in the ${\cal C}_1$ and ${\cal C}_2$ constants 
  (which depend on DM halo mass) of the BPR bias evolution model.}
\end{table}
Note that we do not explicitely correct for possible 
non-linear effects in $\delta$ (although the density field is indeed smoothed on
linear scales  - 8 $h^{-1}$ Mpc); we do however expect that such effects
should be mostly cancelled in the overdensity ratio definition of the bias.

We use the DM halo bias evolution, measured in the two simulations,
for different DM halo mass range subsamples in order to constrain 
the constants of our bias evolution model, ie., ${\cal C}_{1}, {\cal C}_{2}$.
The procedure used is based on a $\chi^2$ minimization of whose details
are presented in BPR and thus will not be repeated here. 
In Fig.A1 we present as points the simulations based values of these
parameters, for both cosmologies used, and as continuous curves their
analytic fits, which are given in eqs.(\ref{eq:C1}) and (\ref{eq:C2}).
The resulting values of the parameters $\alpha_{1,2}$ and
$\beta_{1,2}$ can be found in Table A1.
It is interesting to note that the slope of the functions ${\cal C}_1$
and ${\cal C}_2$ is roughly a constant and independent of cosmology,
with a value $\beta_1 \simeq \beta_2 \simeq 0.35 (\pm 0.06)$.

\begin{figure}
\resizebox{8cm}{8cm}{\includegraphics{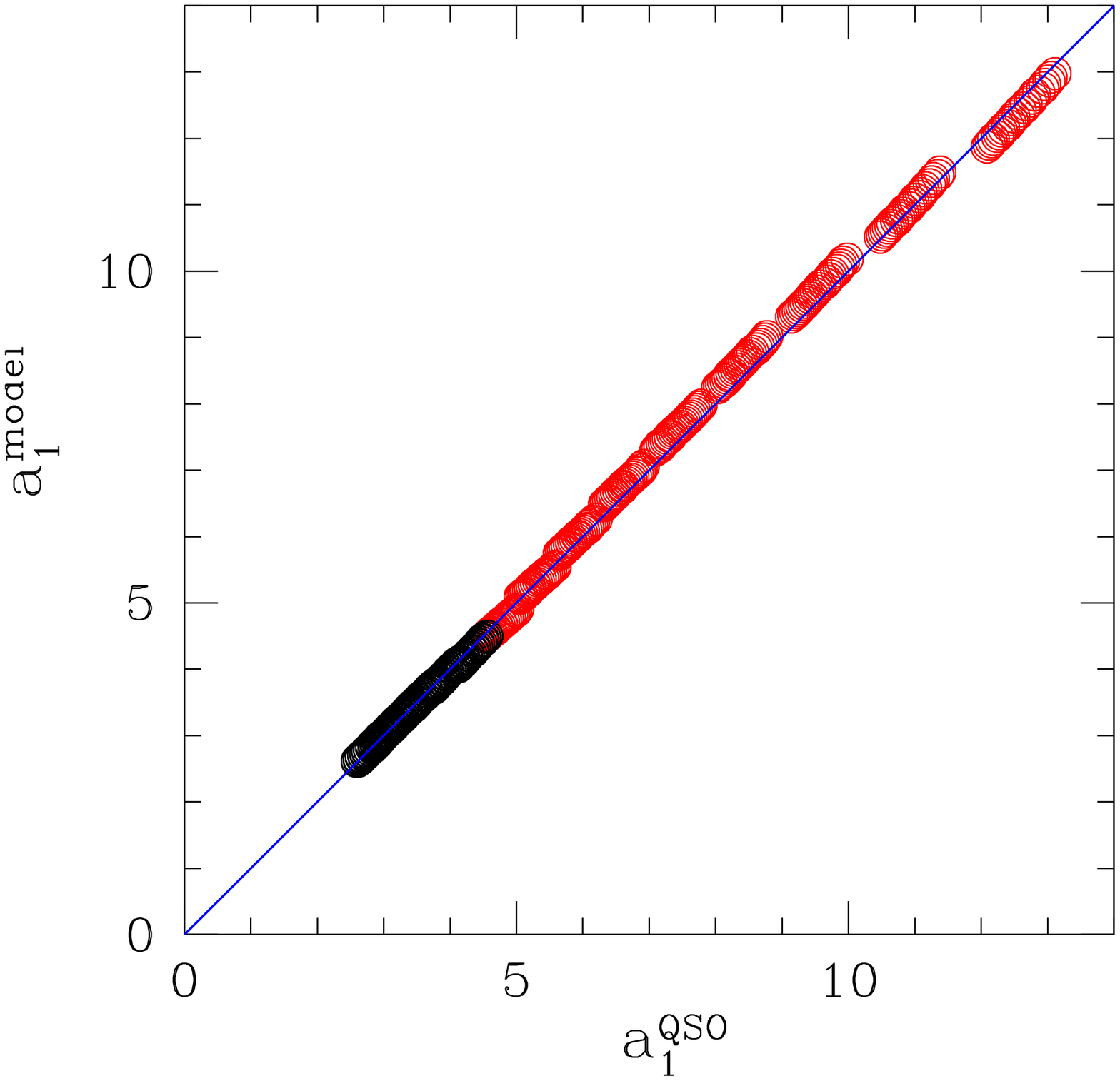}}
\caption{Correlation between (a) the best fitted $\alpha_1$ values of
  the BPR model using the QSO bias data scaled to different
  cosmologies for a  grid of $\Omega_m$ and $\sigma_8$ values and (b) the
  predicted $\alpha_1$ values, based on  eq.(\ref{eq:model}). The red
  points correspond to $\Omega_m\le 0.273$ while the black points to
  $\Omega_m>0.273$.}
\end{figure}

\subsection{Dependence of the BPR model constants on Cosmology}
The dependence of the constants $\alpha_1$ and $\alpha_2$ of the BPR bias model
on the different cosmological parameters is an important prerequisite
for the versatile use of the model in investigating the bias evolution
of different mass tracers and determine the mass of the dark matter halos which
they inhabit.
In Basilakos \& Plionis (2001) we predicted a power law dependence of 
$\alpha_2$ on $\Omega_m$. Indeed, fitting such a
dependence, using the WMAP1 and WMAP7 $\Lambda$CDM simulations, we find:
\be
\alpha_2(\Omega_m) =-0.41 \left(\frac{0.273}{\Omega_m}\right)^{n} \;\;{\rm with}\;\;
n\simeq 2.8/2
\ee
consistent with the value $n=3/2$ anticipated in Basilakos \& Plionis
(2001).

Now, in order to investigate the dependence on different cosmological
parameters of the parameter $a_1$, we have used the optical QSO 
data and the procedure outlined in section 2.2 to scale
the QSO bias data to different flat cosmologies, using a grid of $\Omega_m$
and $\sigma_8$ values. The grid was defined as follows: $\Omega_m \in
[0.18, 0.5]$ and $\sigma_8 \in [0.7, 0.94]$, both in steps
of 0.01. We then minimize the BPR bias evolution 
model to the scaled bias data to different cosmologies QSO, finally
providing for each pair of ($\Omega_m, \sigma_8$) values the best
fitted $\alpha_1$ and $M_h$ values.

Then using a trial and error approach to select the best functional 
dependence of the derived $\alpha_1(\Omega_m, \sigma_8)$ values to the relevant
cosmological parameters, we find a best fit model of the form:
\be\label{eq:model}
\alpha_1(\Omega_m, \sigma_8)\simeq 4.53 \left(\frac{0.81}{\sigma_{8}}\right)^{\kappa_1}
\exp{\left[\kappa_2(\Omega_m-0.273)\right]}$$
\ee
with 
\be
(\kappa_1, \kappa_2)= \left\{\begin{array}{ll}
(12.15, 0.30) & \Omega_m\le 0.273 \\
(8.70, 0.37) & \Omega_m> 0.273 
\end{array}
\right.
\ee
In Fig. A2 we correlate the derived $\alpha_1(\Omega_m, \sigma_8)$
values, resulting from fitting the BPR bias evolution model to the
scaled QSO bias data, to
those predicted by the model of eq.(\ref{eq:model}). It is evident
that the correspondence is excellent, indicating that indeed the above
estimated cosmological dependence of $\alpha_{1}$ is the indicated one.


\end{document}